\documentclass[twocolumn,showpacs,aps,prl,superscriptaddress]{revtex4}
\usepackage{graphicx}
\usepackage{dcolumn}
\usepackage{amsmath}
\usepackage{epsfig}

\RequirePackage{xspace}

\usepackage{relsize}
\def\babar{\mbox{\slshape B\kern-0.1em{\smaller A}\kern-0.1em
    B\kern-0.1em{\smaller A\kern-0.2em R}}}

\def\epem       {\ensuremath{e^+e^-}\xspace}

\def\qqbar {\ensuremath{q\overline q}\xspace}

\def\piz   {\ensuremath{\pi^0}\xspace}

\def\pip   {\ensuremath{\pi^+}\xspace}

\def\pipi  {\ensuremath{\pi^+\pi^-}\xspace}

\def\Kbar  {\kern 0.2em\overline{\kern -0.2em K}{}\xspace}

\def\Kz    {\ensuremath{K^0}\xspace}
\def\Kzb   {\ensuremath{\Kbar^0}\xspace}
\def\KzKzb {\ensuremath{\Kz \kern -0.16em \Kzb}\xspace}
\def\Kp    {\ensuremath{K^+}\xspace}
\def\Km    {\ensuremath{K^-}\xspace}

\def\KpKm  {\ensuremath{\Kp \kern -0.16em \Km}\xspace}

\def\Dbar    {\kern 0.2em\overline{\kern -0.2em D}{}\xspace}

\def\Dz      {\ensuremath{D^0}\xspace}
\def\Dzb     {\ensuremath{\Dbar^0}\xspace}
\def\DzDzb   {\ensuremath{\Dz {\kern -0.16em \Dzb}}\xspace}
\def\Dp      {\ensuremath{D^+}\xspace}
\def\Dm      {\ensuremath{D^-}\xspace}

\def\DpDm    {\ensuremath{\Dp {\kern -0.16em \Dm}}\xspace}

\def\Dstarp  {\ensuremath{D^{*+}}\xspace}

\def\B       {\ensuremath{B}\xspace}
\def\Bbar    {\kern 0.18em\overline{\kern -0.18em B}{}\xspace}

\def\BB      {\ensuremath{B\Bbar}\xspace} 
\def\Bz      {\ensuremath{B^0}\xspace}
\def\Bzb     {\ensuremath{\Bbar^0}\xspace}
\def\BzBzb   {\ensuremath{\Bz {\kern -0.16em \Bzb}}\xspace}
\def\Bu      {\ensuremath{B^+}\xspace}
\def\Bub     {\ensuremath{B^-}\xspace}
\def\Bp      {\ensuremath{\Bu}\xspace}
\def\Bm      {\ensuremath{\Bub}\xspace}
\def\Bpm     {\ensuremath{B^\pm}\xspace}

\def\BpBm    {\ensuremath{\Bu {\kern -0.16em \Bub}}\xspace}

\def\BorBbar    {\optbar{\kern +0.13em B}{}\xspace}
\def\DorDbar    {\optbar{\kern +0.13em D}{}\xspace}
\def\KorKbar    {\optbar{\kern +0.13em K}{}\xspace}

\mathchardef\Upsilon="7107
\def\Y#1S{\ensuremath{\Upsilon{(#1S)}}\xspace}

\def\FourS {\Y4S}

\mathchardef\Deltares="7101
\mathchardef\Xi="7104
\mathchardef\Lambda="7103
\mathchardef\Sigma="7106
\mathchardef\Omega="710A

\def\Deltabar{\kern 0.25em\overline{\kern -0.25em \Deltares}{}\xspace}
\def\Lbar{\kern 0.2em\overline{\kern -0.2em\Lambda\kern 0.05em}\kern-0.05em{}\xspace}
\def\Sigbar{\kern 0.2em\overline{\kern -0.2em \Sigma}{}\xspace}
\def\Xibar{\kern 0.2em\overline{\kern -0.2em \Xi}{}\xspace}
\def\Obar{\kern 0.2em\overline{\kern -0.2em \Omega}{}\xspace}
\def\Nbar{\kern 0.2em\overline{\kern -0.2em N}{}\xspace}
\def\Xb{\kern 0.2em\overline{\kern -0.2em X}{}\xspace}

\def\BR         {{\ensuremath{\cal B}\xspace}}

\def\Bztopipi   {\ensuremath{\Bz \to \pipi}\xspace}

\def\mes        {\mbox{$m_{\rm ES}$}\xspace}

\newcommand{\tev}{\ensuremath{\mathrm{\,Te\kern -0.1em V}}\xspace}
\newcommand{\gev}{\ensuremath{\mathrm{\,Ge\kern -0.1em V}}\xspace}
\newcommand{\mev}{\ensuremath{\mathrm{\,Me\kern -0.1em V}}\xspace}
\newcommand{\kev}{\ensuremath{\mathrm{\,ke\kern -0.1em V}}\xspace}
\newcommand{\ev}{\ensuremath{\mathrm{\,e\kern -0.1em V}}\xspace}
\newcommand{\gevc}{\ensuremath{{\mathrm{\,Ge\kern -0.1em V\!/}c}}\xspace}
\newcommand{\mevc}{\ensuremath{{\mathrm{\,Me\kern -0.1em V\!/}c}}\xspace}
\newcommand{\gevcc}{\ensuremath{{\mathrm{\,Ge\kern -0.1em V\!/}c^2}}\xspace}
\newcommand{\mevcc}{\ensuremath{{\mathrm{\,Me\kern -0.1em V\!/}c^2}}\xspace}

\def\invfb   {\ensuremath{\mbox{\,fb}^{-1}}\xspace}

\def\mus  {\ensuremath{\rm \,\mus}\xspace}

\def\mus        {\ensuremath{\,\mu{\rm s}}\xspace}    

\def\to                 {\ensuremath{\rightarrow}\xspace}

\def\pep2{PEP-II}

\def\gsim{{~\raise.15em\hbox{$>$}\kern-.85em
          \lower.35em\hbox{$\sim$}~}\xspace}
\def\lsim{{~\raise.15em\hbox{$<$}\kern-.85em
          \lower.35em\hbox{$\sim$}~}\xspace}

\def\CP                {\ensuremath{C\!P}\xspace}

\xspace

\newcommand{\epjBase}        {Eur.\ Phys.\ Jour.\xspace}
\newcommand{\jprlBase}       {Phys.\ Rev.\ Lett.\xspace}
\newcommand{\jprBase}        {Phys.\ Rev.\xspace}
\newcommand{\jplBase}        {Phys.\ Lett.\xspace}

\newcommand{\zpBase}         {Z.\ Phys.\xspace}

\newcommand{\epjc}      [1]  {\epjBase\ C~{\bf #1}}

\newcommand{\plb}       [1]  {\jplBase\ B~{\bf #1}}

\newcommand{\jprl}      [1]  {\jprlBase\ {\bf #1}}
\newcommand{\jprd}      [1]  {\jprBase\ D~{\bf #1}}

\newcommand{\progtp}    [1]  {{Prog.\ Theor.\ Phys.\ {\bf #1}}}

\newcommand{\zpc}       [1]  {\zpBase\ C~{\bf #1}}

\newcommand{\hepex}     [1]  {hep-ex/{#1}}
\newcommand{\hepph}     [1]  {hep-ph/{#1}}

\def\jetset74   {\mbox{\tt Jetset \hspace{-0.5em}7.\hspace{-0.2em}4}\xspace}

\newcommand{\etal}{{\it et al.}}
\def\Btopipi   {\ensuremath{\B \to \pi\pi}\xspace}

\def\Bptopipiz   {\ensuremath{\B^+ \to \pi^+\piz}\xspace}
\def\Btokpi   {\ensuremath{\B \to K\pi}\xspace}
\def\Btokppim   {\ensuremath{\Bz \to K^+\pi^-}\xspace}

\def\Bptokpiz   {\ensuremath{\B^+ \to K^+\piz}\xspace}

\def\Bptohpiz   {\ensuremath{\B^+ \to \hp\piz}\xspace}
\def\Bztopizpiz   {\ensuremath{\Bz \to \piz\piz}\xspace}

\def\Btorhoppim   {\ensuremath{\Bz \to \rho^+\pi^-}\xspace}

\def\Btorhomkp   {\ensuremath{\Bz \to \rho^- K^+}\xspace}

\def\Bptorhoppiz   {\ensuremath{\B^+ \to \rho^+\piz}\xspace}

\def\Btag {\ensuremath{B_{\rm tag}}}

\def\hp    {\ensuremath{h^+}\xspace}
\def\acp {\ensuremath{{\cal A}}\xspace}
\def\acppipiz {\ensuremath{{\cal A}_{\pi^+\piz}}\xspace}
\def\acpkpiz {\ensuremath{{\cal A}_{K^+\piz}}\xspace}

\def\alphaeff {\ensuremath{\alpha_{\rm eff}}\xspace}
\def\de {\ensuremath{\Delta E}\xspace}
\def\cossph   {\ensuremath{|\cos{\theta_{\scriptscriptstyle S}}|\;}\xspace}
\def\fish    {\ensuremath{F}\xspace}

\def\deltapipi {\ensuremath{\delta_{\pi\pi}}}
\def\sss{\scriptscriptstyle}
\def\barpd{{\raise.35ex\hbox
{${\sss (}$}}--{\raise.35ex\hbox{${\sss )}$}}}
\def\BorBbar{\hbox{$B^{0}$\kern-1.25em\raise1.5ex\hbox{\barpd}}}
\long\def\inst#1{\par\nobreak\kern 4pt\nobreak
    {\it #1}\par\vskip 10pt plus 3pt minus 3pt}

\newcommand{\BABARPubYear}    {04}
\newcommand{\BABARPubNumber}  {042}
\newcommand{\SLACPubNumber} {10902}

\begin{document}

\noindent
\babar-PUB-\BABARPubYear/\BABARPubNumber \\
SLAC-PUB-\SLACPubNumber \\

\title{
{\large \bf
Branching Fractions and {\boldmath \CP} Asymmetries in
{\boldmath \Bztopizpiz, \Bptopipiz and\\
\Bptokpiz} Decays and Isospin Analysis of the {\boldmath \Btopipi} System
}}

%
\author{B.~Aubert}
\author{R.~Barate}
\author{D.~Boutigny}
\author{F.~Couderc}
\author{Y.~Karyotakis}
\author{J.~P.~Lees}
\author{V.~Poireau}
\author{V.~Tisserand}
\author{A.~Zghiche}
\affiliation{Laboratoire de Physique des Particules, F-74941 Annecy-le-Vieux, France }
\author{E.~Grauges-Pous}
\affiliation{Universitad Autonoma de Barcelona, E-08193 Bellaterra, Barcelona, Spain }
% ...
\collaboration{The \babar\ Collaboration}
\noaffiliation
\date{\today}

\begin{abstract}
Based on a sample of 227 million
\BB\ pairs collected by the \babar\ detector at the \pep2\
asymmetric-energy $B$ Factory at SLAC,
we measure the branching fraction
$\BR(\Bztopizpiz) = (1.17 \pm 0.32 \pm 0.10) \times 10^{-6}$, 
and the asymmetry 
$C_{\piz\piz}= -0.12 \pm 0.56 \pm 0.06 $.
The \Bztopizpiz signal has a significance of $5.0\sigma$.
We also measure
${\BR}(\Bptopipiz)   = (5.8 \pm 0.6 \pm 0.4) \times 10^{-6}$,
${\BR}(\Bptokpiz)   = (12.0 \pm 0.7 \pm 0.6) \times 10^{-6}$, 
and the charge asymmetries
$ \acppipiz   = -0.01\pm 0.10\pm 0.02 $ and
$  \acpkpiz   = 0.06 \pm 0.06 \pm 0.01$.
Using isospin relations we
find an upper bound on the angle difference  $\left|\alpha - \alpha_{\rm eff} \right|$
of $35^{\rm o}$ at the \mbox{$90\%$ C.L.}
\end{abstract}


\maketitle

In the Standard Model (SM), 
the Cabibbo-Kobayashi-Maskawa (CKM) matrix $V_{qq^\prime}$~\cite{ckmref}
describes the charged-current couplings in the quark sector. 
The Unitarity Triangle is a useful representation of relations between
CKM matrix elements, and 
measurements of its sides and angles 
provide a stringent test of the SM.
Following the success in measuring the CKM angle $\beta$~\cite{sin2beta},
an important challenge for the $B$ Factories is the determination of the remaining angles.
The extraction of the CKM angle
$\alpha \equiv \arg\left[-V_{\rm td}^{}V_{\rm tb}^{*}/V_{\rm ud}^{}V_{\rm ub}^{*}\right]$ 
from the time-dependent \CP-violating asymmetry in the
\Bztopipi decay mode~\cite{sin2alpha} is complicated by the interference of 
competing amplitudes (``tree'' and ``penguin'') with different weak phases.  The difference between
$\alpha$ and \alphaeff,
where \alphaeff is
derived from the time-dependent \Bztopipi \CP asymmetry,
may be evaluated using the
isospin-related decays \Bztopizpiz and
\Bptopipiz~\cite{Isospin}. Here and throughout this Letter, charge
conjugate reactions are included implicitly.
For \Bztopizpiz
the asymmetry
may deviate from zero  if the tree
and penguin amplitudes have different weak and strong phases.
In the SM  the
decay \Bptopipiz is governed by a pure tree amplitude since penguin diagrams cannot
contribute to the $I=2$ final state; as a result no charge
asymmetry is expected. 
The \Btokpi system is a rich source of information on the understanding of
\CP violation, as has been illustrated by the recent observation of
direct \CP asymmetry in \Btokppim decays~\cite{AKpi}. Both the rate and
asymmetry of the \Bptokpiz decay may be used to extract constraints on
penguin contributions to the \Btokpi amplitudes~\cite{SumRules}.

In this Letter, we report
a constraint on \mbox{$\deltapipi \equiv \alphaeff - \alpha$}, 
using the measurement of the asymmetry $C_{\piz\piz}$ and
updated measurements of the
branching fractions for \Bztopizpiz and \Bptopipiz
and the charge asymmetry \acppipiz. 
We also measure the branching fraction for the \Bptokpiz decay and its charge asymmetry \acpkpiz. 
The asymmetry
$C_{\piz\piz}$ is defined as
$(\left|A_{00}\right|^2 - \left|\overline{A}_{00}\right|^2)/(\left|A_{00}\right|^2 + \left|\overline{A}_{00}\right|^2)$,
where $A_{00}$ ($\overline{A}_{00}$) is the
{\ensuremath{\Bz (\Bzb) \to \piz\piz}\xspace}
decay amplitude.
For \Bpm modes, the \CP-violating charge asymmetry is defined as
$\acp =  (|\overline{A}|^2 - |A|^2 )/(|\overline{A}|^2 +  |A|^2 )$,
where $A$ ($\overline{A}$) is the \Bp (\Bm) decay amplitude.
This study is  based on $227 \times 10^{6}$ $\FourS\to\BB$ decays 
(on-resonance), collected with the \babar\ detector. We also use
$16\invfb$ of data recorded $40\mev$ below
the \BB\ production threshold (off-resonance).

The \babar\ detector is described in Ref.~\cite{babarnim}.
The primary components used in this analysis are a
tracking system consisting of a five-layer silicon
vertex tracker (SVT) and a 40-layer drift chamber (DCH) surrounded
by a $1.5$ T solenoidal magnet, an electromagnetic calorimeter
(EMC) comprising $6580$ CsI(Tl) crystals, and a 
ring imaging Cherenkov counter (DIRC).

Candidate \piz\ mesons are reconstructed as pairs of \mbox{photons}, 
spatially
separated in the EMC,  
with an invariant mass 
$m_{\gamma \gamma}$
satisfying $110<m_{\gamma \gamma}<160$\mevcc.
The mass resolution is 8~\mevcc for high energy 
(above $2 \gev$) \piz's~\cite{babarnim}.
Photon candidates are required to be consistent with 
the expected lateral shower shape, not to be matched to a track, and to have a
minimum energy of 30\mev. 
To reduce the background from false \piz candidates, the
angle $\theta_{\gamma}$ between the photon momentum vector in the \piz rest frame and
the \piz 
flight direction
is required to satisfy
$|\cos{\theta_{\gamma}}| < 0.95$.
Candidate tracks are required to be within the
tracking fiducial volume, to originate from the interaction point,  
to consist of at least 12 DCH hits, and to be associated with at least 6 Cherenkov
photons in the DIRC.

\B meson candidates are reconstructed by combining a
\piz with a charged pion or kaon (\hp) or by combining two \piz mesons.
Two variables, used to isolate the \Bztopizpiz and \Bptohpiz signal events, take
advantage of the kinematic constraints of \B mesons produced at the
\FourS. The first is the beam-energy-substituted mass $\mes$ $=$ $\sqrt{
  (s/2 + {\bf p}_{i}\cdot{\bf p}_{B})^{2}/E_{i}^{2}- {\bf
    p}^{2}_{B}}$, 
where 
$(E_{i},{\bf p}_{i})$ is the four-momentum of the initial \epem system,
${\bf p}_{B}$ is the \B candidate momentum, both measured in
the laboratory frame,
and $\sqrt{s}$ is the \epem center-of-mass (CM) energy.
The second variable is \de 
$ = E_{B} -
\sqrt{s}/2$, where $E_{B}$ is the \B candidate energy in the CM frame.
The \de resolution for signal is approximately 80~\mev for \Bztopizpiz,
and 40~\mev for \Bptohpiz.

The primary source of background is $\epem \to \qqbar \;(q =
u,d,s,c)$ events where a \piz
or \hp from each jet randomly combine to mimic a \B decay.
This jet-like \qqbar background is suppressed by
requiring that the angle $\theta_{\scriptscriptstyle S}$ between the
sphericity axis of the \B candidate and that of the
remaining tracks and photons in the event, in the CM frame, satisfy
$\cossph < 0.7 \,(0.8)$
for \Bztopizpiz (\Bptohpiz).
The other sources of background are \B decays to
final states containing one vector meson and one pseudoscalar meson, 
where one pion is produced almost at rest in the \B rest frame and
the remaining decay products match the kinematics of a
\Bztopizpiz or \Bptohpiz decay.

For the \Bztopizpiz analysis we restrict the \mbox{$\mes$-$\de$} plane to the region with
$\mes>5.2$\gevcc and $|\de|<0.4$ \gev. For the on-resonance sample we
define the signal region as 
the band in the plane with $|\de|<0.2$\gev and the
sideband region as the rest of the plane 
excluding the region which is also populated with \Bptorhoppiz events.
The entire plane for the off-resonance data and the sideband region for the
on-resonance data 
are kept in the fit in order to  constrain the \qqbar background
parameters.
\Bptohpiz candidates are selected in the region with $\mes>5.22$\gevcc and  $-0.11<\de<0.15$\gev. 

For \Bztopizpiz candidates, the other tracks and clusters in the event are used to determine whether the other
$B$ meson (\Btag) decays as a $\Bz$ or $\Bzb$ (flavor tag).
We use a multivariate technique~\cite{sin2beta04} to determine the flavor of
the $\Btag$ meson.  
Events are assigned to one of several mutually exclusive categories
based on the estimated mistag probability and on the source of tagging information.

The number of signal \B candidates is determined with an extended, unbinned
maximum-likelihood fit.  
The probability density function (PDF) ${\cal P}_i\left(\vec{x}_j;
  \vec{\alpha}_i\right)$
for a signal or background hypothesis
is the product of PDFs for the
variables $\vec{x}_j$ given the set of parameters $\vec{\alpha}_i$.
The likelihood function is a product over the $N$ events of the
$M$ signal and background hypotheses:
\begin{equation}
{\cal L}= \exp\left(-\sum_{k=1}^M n_k\right)\,
\prod_{j=1}^N \left[\sum_{i=1}^M c_{ij} {\cal P}_i\left(\vec{x}_j;
\vec{\alpha}_i\right)
\right]\, .
\end{equation}

For \Bztopizpiz the coefficients $c_{ij}$ are defined as 
$c_{ij} = \frac{1}{2}(1-s_j A_i) n_i$, where $s_j$ refers to the sign of the flavor tag of the other \B in  the
event $j$ and is zero for untagged events. The fit parameters 
$n_i$ and $A_i$ are the number of events and raw asymmetry 
for \Bztopizpiz signal, \Bptorhoppiz background,
and continuum background components.  The average of branching fraction measurements~\cite{rhopi0}
is used to fix $n(\Bptorhoppiz)$ to $32\pm 6$. The raw asymmetry for signal
is $(1-2\chi)(1-2\omega)C_{\piz\piz}$,
where $\chi = 0.186\pm 0.004$~\cite{PDG2004} is the neutral \B mixing probability, 
and $\omega$ is the mistag probability.

For \Bptohpiz the probability coefficients are $c_{ij} = \frac{1}{2}(1-q_j\acp_i)n_i$, where $q_j$ is the
charge of the track $h$ in the event $j$.   The fit parameters $n_i$ and $\acp_i$ are the number
of events and asymmetry
for \Bptopipiz and \Bptokpiz
signal, continuum, and \B background  components.
The \B background yields are fixed to the
expected number of events using the current world averages of branching ratios~\cite{rhopirhok},
which are $18\pm 4$ for \Btorhoppim and \Bptorhoppiz combined, and $3\pm 1$
events for \Btorhomkp.
Uncertainties on these numbers
are dominated by the uncertainty on selection efficiencies,
due to the sensitivity to the tight requirement in \de.

The variables $\vec{x}_j$ used for \Bztopizpiz are \mes, \de, and
a Fisher discriminant \fish.
The Fisher
discriminant is an optimized linear combination of $\sum_i
p_i$ and $\sum_i p_i \cos^2{\theta_i}$,
where $p_{i}$ is the momentum and $\theta_{i}$ is the angle with
respect to the thrust axis of the \B candidate, both in the CM frame,
for all tracks and neutral clusters not used to reconstruct the \B
meson.
For both the  \Bztopizpiz signal and the \Bptorhoppiz
background the \mes and \de variables are correlated and therefore a 
two-dimensional PDF  from a smoothed, simulated distribution is used.
For the continuum background, the \mes distribution is modeled as a threshold function~\cite{Argus}, 
and the \de distribution as a second-order polynomial.
The PDF for the \fish variable is modeled as a parametric step  
function (PSF)~\cite{previous} for all event components.  
A PSF is a variable width binned distribution whose parameters are the
heights of each bin.
The limits of the ten bins \fish PSF are
chosen so that each bin contains 10\% of the signal sample.
For \Bztopizpiz and \Bptorhoppiz the \fish PSF parameters
are correlated with the flavor tagging, and the PSF parameters are
different for each tagging category. 
Simulated events are used to determine the PSF
distributions for both \Bztopizpiz and \Bptorhoppiz. 
For \qqbar background, the \fish PSF parameters are free in
the fit.

An additional discriminating variable for \Bptohpiz is the
Cherenkov angle $\theta_c$ of the \hp track.
The PDF parameters for \mes, \de, $\theta_c$, and \fish for the background are
determined using the data, while the PDFs for signal are found from a
combination of simulated events and data.  
The \mes and \de distributions for \qqbar events are treated as in
the \Bztopizpiz case, with parameters allowed to vary freely in the fit.
For the signal, the \mes and \de
distributions are both modeled as a Gaussian distribution with a low-side power law tail
 whose parameters are determined from simulation.
The means of the Gaussian components are determined from the fit
to the \Bptohpiz sample and  their values used to tune the \piz energy scale in the 
\Bztopizpiz analysis. The mean of \de
for the \Bptokpiz mode is a function of the kaon laboratory momentum,
since a pion mass hypothesis is used.
The distribution of \fish is
modeled as a Gaussian function with an asymmetric variance for the signal, 
whose parameters are obtained from simulation,
 and as  a double Gaussian for 
the continuum background, whose parameters 
are determined in the likelihood fit. The difference of the measured and
expected values of $\theta_c$ for the pion or kaon hypothesis,
divided by the uncertainty on $\theta_c$, is modeled as a double
Gaussian function, whose parameters are obtained from a 
control sample of kaon and pion tracks, from $\Dstarp \to
\Dz \pip \, , \Dz \to \Km \pip$ decays.

\begin{table*}[htbp]
\begin{center}
\caption{ The results for the modes \Bztopizpiz and \Bptohpiz are summarized.  For each mode, the
sample size $N$, number of signal events $N_S$, total detection efficiency $\varepsilon$,
branching fraction \BR, asymmetry ${\cal A}$ or $C_{\piz\piz}$, and the $90\%$ 
confidence interval for the asymmetry are shown.
For $C_{\piz\piz}$ the confidence interval is obtained inferring minimum coverage inside 
the physical region $[-1, 1]$.
The first errors are statistical, the second systematic, with the exception of $\varepsilon$ whose
error is purely systematic.}
\label{table:summary}
\begin{tabular}{l|cccccc}
Mode        & $N$    & $N_{S}$    & $\varepsilon$ (\%) & \BR($10^{-6}$)       & Asymmetry      & $(90\%$  ${\rm C.L.})$  
\\
\hline\\[-2.5ex] 
\Bztopizpiz &   $8153$  &   $61\pm 17$ & $23.5\pm 1.4$   &  $1.17\pm 0.32\pm 0.10$ &  $-0.12\pm 0.56\pm 0.06$ &  $[-0.88, 0.64]$
\\
\Bptopipiz   &  $29950$  & $379\pm 41$ &  $28.7\pm 1.1$ &  $5.8\pm 0.6\pm 0.4$ & $-0.01\pm 0.10\pm 0.02$ &  $[-0.19 , 0.21]$ 
\\
\Bptokpiz    &  $13165$  &  $682\pm 39$ & $25.0\pm 1.0$    & $12.0\pm 0.7\pm 0.6$ & $ \ \ 0.06\pm 0.06\pm 0.01$ &  $[-0.06 , 0.18]$
\\
\end{tabular}
\end{center}
\vspace{-3mm}
\end{table*}

The result of the maximum-likelihood fit for \Bztopizpiz is
$n(\Bztopizpiz) = 61\pm 17$ (see Table~\ref{table:summary}),
with a corresponding statistical significance of $5.2\sigma$. 
The asymmetry is $C_{\piz\piz} = -0.12\pm 0.56$.
\begin{figure}[htbp]
\begin{center}
  \includegraphics[width=0.486\linewidth]{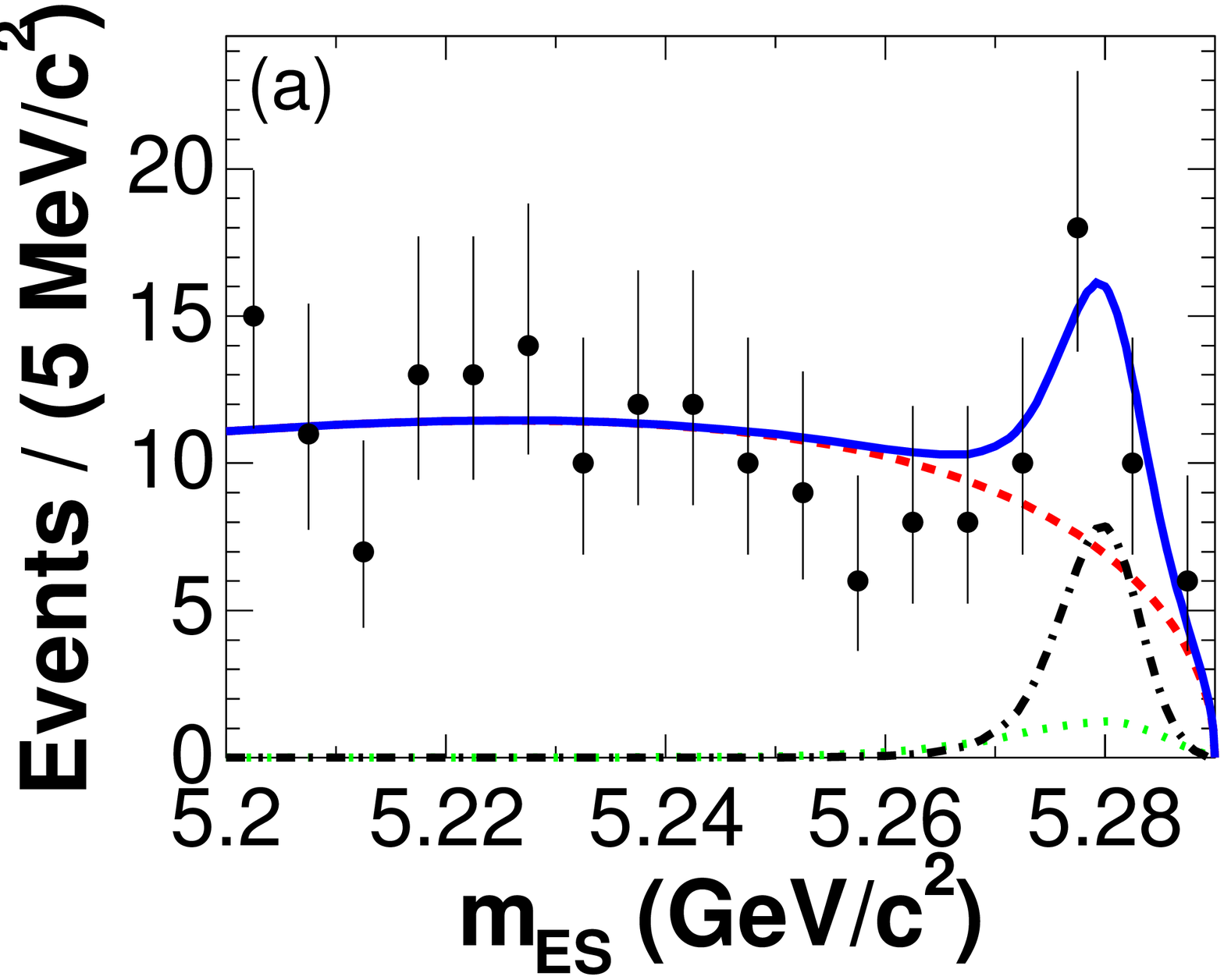}
\hfill
  \includegraphics[width=0.486\linewidth]{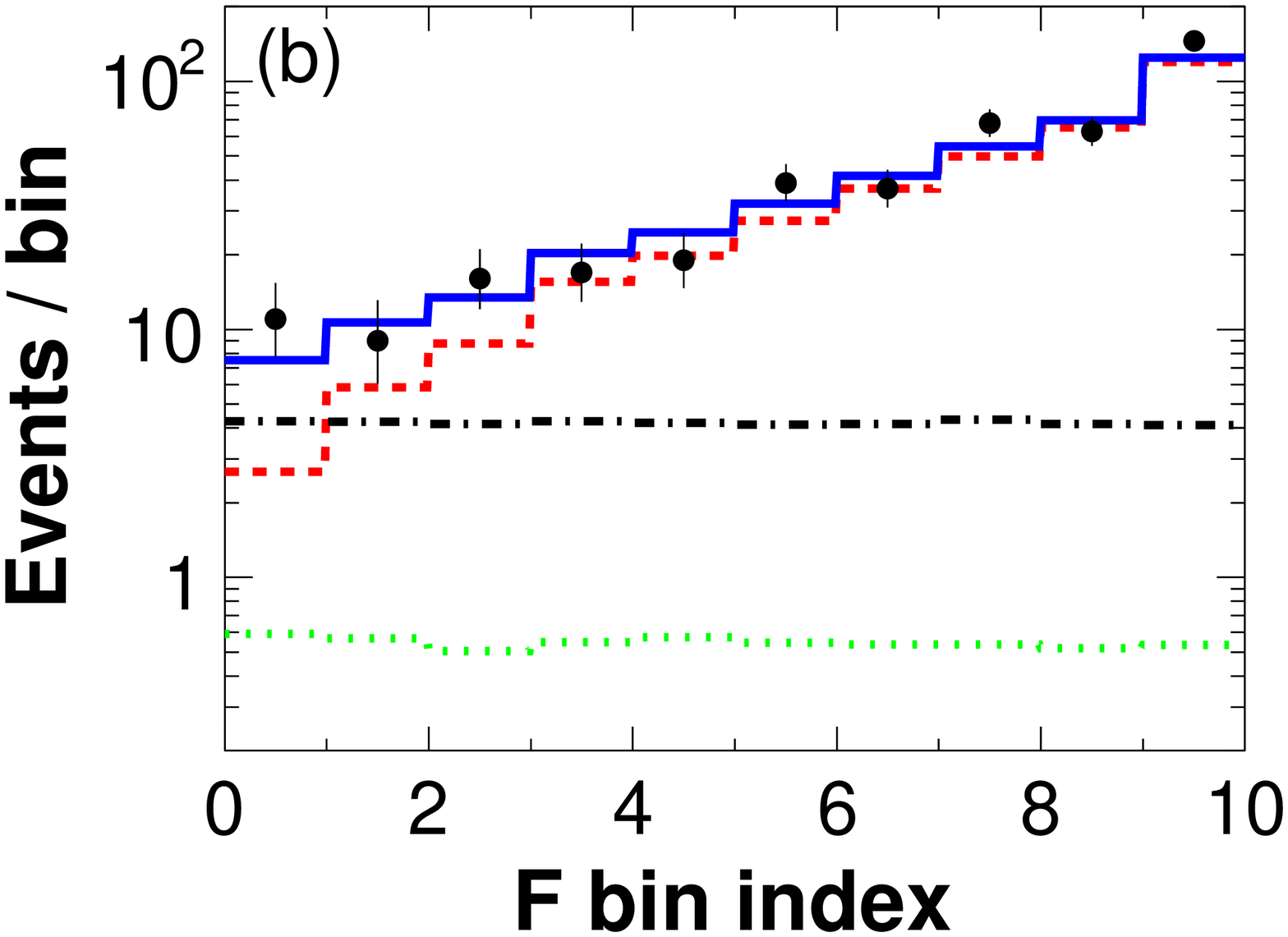}
\end{center}
\vspace{-0.5cm}
\caption{Distributions and PDF projections for \Bztopizpiz.
Shown are \mes (a) and \fish (b) for candidates that satisfy an optimized
requirement on the signal probability, based on
all variables except the one being plotted. PDF projections are shown as a dashed
line for \qqbar background, a dotted line for \B background, and a
dashed-dotted line for signal.}
\label{fig:pi0pi0only}
\end{figure}
Shown in Fig.~\ref{fig:pi0pi0only} are
distributions of
\mes and \fish, for signal-enriched samples of \Bztopizpiz candidates.
The projections contain
25\% and 68\%
of the signal,
14\% and 17\%
of the $\rho^+\piz$
background, and
2.2\% and 4.4\%
of the continuum background,
for \mes and \fish respectively.
 
With changes in the analysis technique to measure
the \CP asymmetry, we now find $44\pm 13$ signal events in the first 123 million \BB\ events, compared
to $46\pm 13$ found in Ref.~\cite{previous}. The additional 104 million \BB\ events
dataset has a signal of $17\pm 11$. The signal rates in these two subsets agree
at the $1.3\sigma$ level. 
This result also reflects an improved understanding of high energy 
\piz detection efficiency.
Using a sample of \piz mesons from
$\tau^+\to \pi^+ \piz \nu_{\tau}$ decays we apply a \piz efficiency
correction of $0.99 \pm 0.03$ to our GEANT simulation,
compared to a correction of $0.88 \pm 0.08$ applied in
Ref.~\cite{previous}.  

\begin{figure}[htbp]
\begin{center}
  \includegraphics[width=0.486\linewidth]{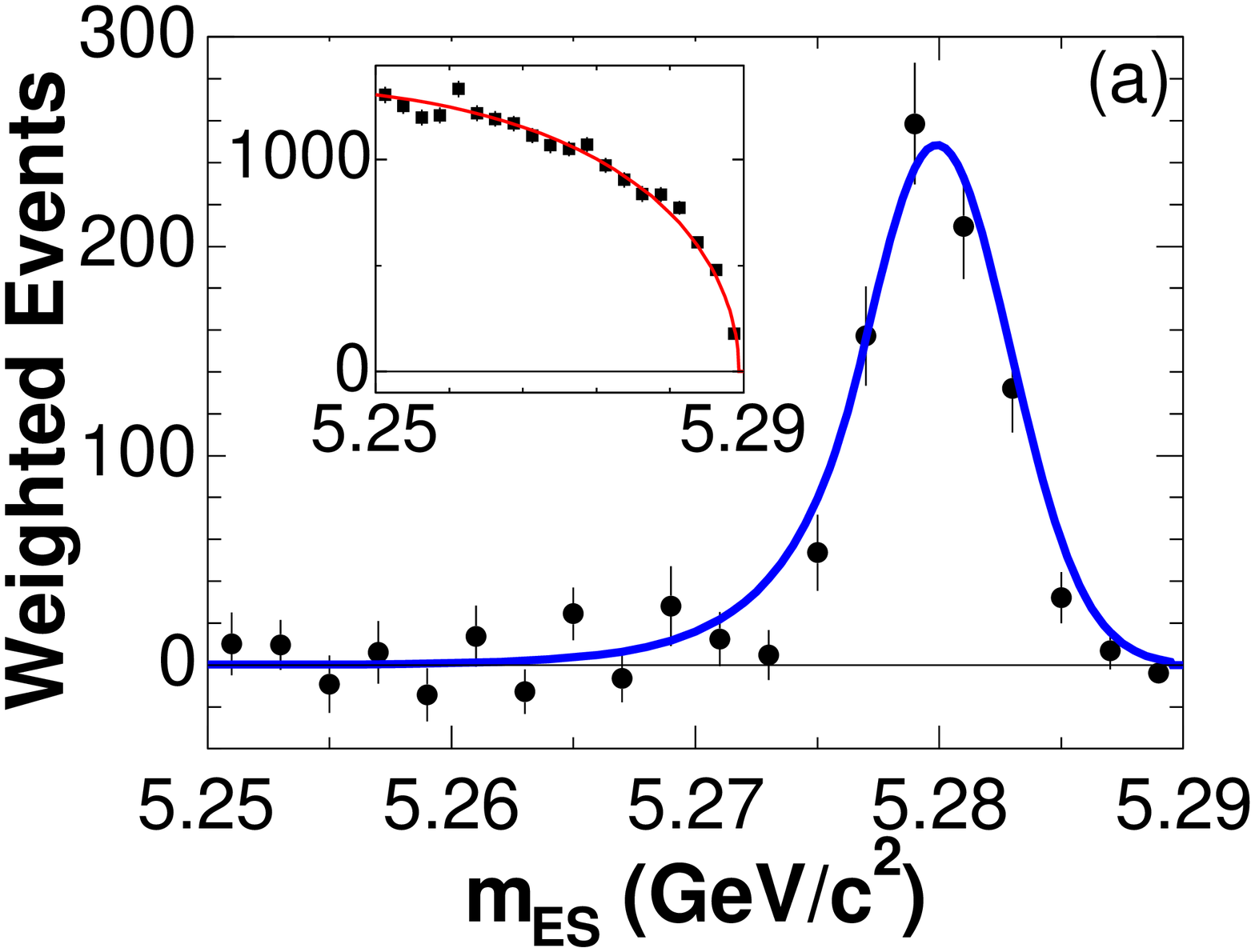}
\hfill
  \includegraphics[width=0.486\linewidth]{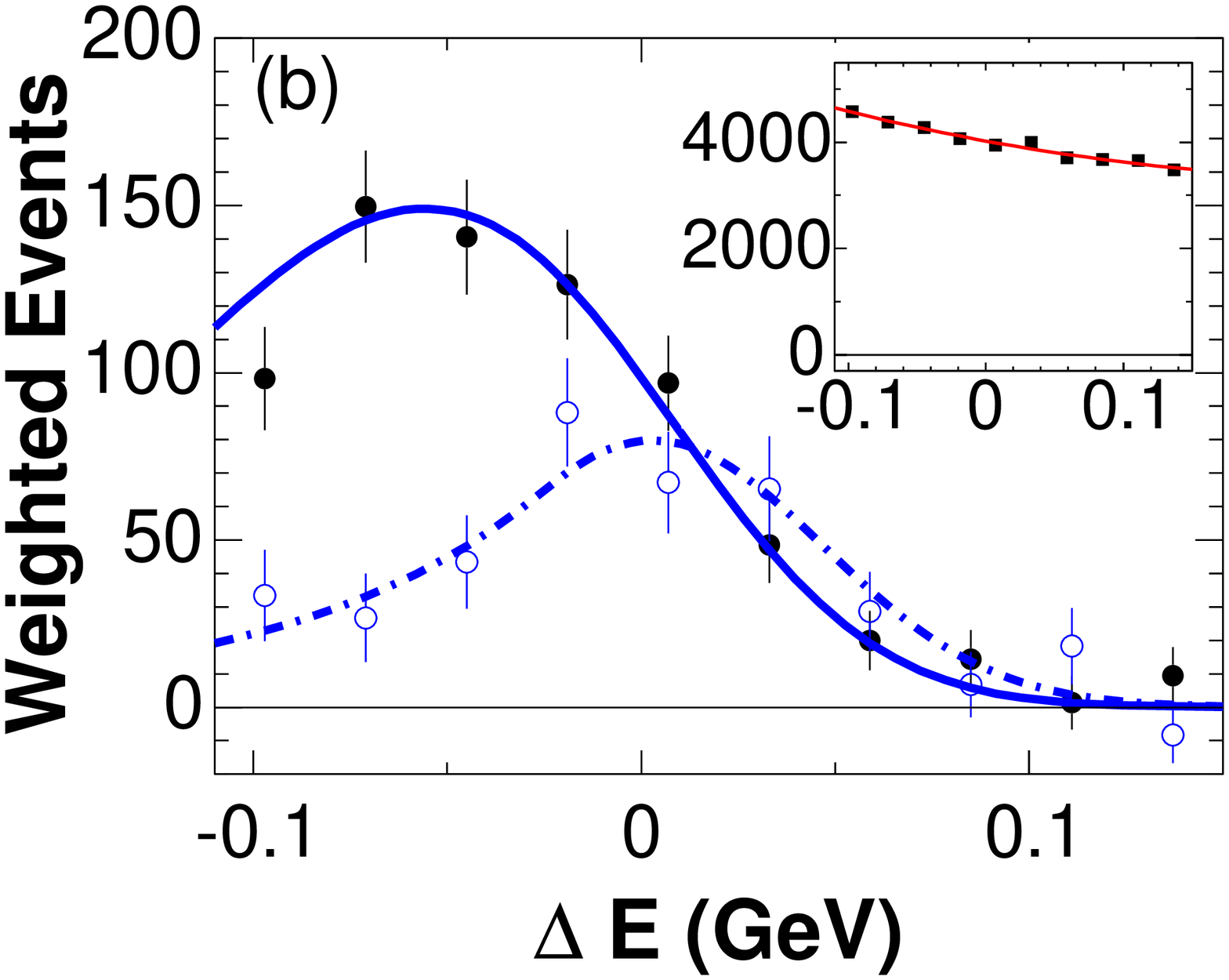}
\end{center}
\vspace{-0.5cm}
\caption{
Distributions and PDF projections for \Bptohpiz, using the method
described in the text. 
For \mes (a) the signal distributions are combined,
while for \de (b) the signal \Bptopipiz (open circles and dashed-dotted curve) and \Bptokpiz (solid circles and curve)
are shown separately. 
The insets show the combined background components.}
\label{fig:hpi0only}
\end{figure}

For \Bptohpiz the likelihood fit results are summarized
in Table~\ref{table:summary}.
Using the event-weighting technique described in Ref.~\cite{splots} we
show signal and background projections in Fig.~\ref{fig:hpi0only}.
For each event, a weight to be signal or
background is assigned based on a fit performed without the specific variable that is plotted.
The resulting distributions are normalized to the event yields,
and are compared to the PDFs used in the full fit.

Systematic uncertainties on the event yields and \CP asymmetries
are evaluated on data control samples, or by
varying the fixed parameters and refitting the data.
In order of decreasing importance,
the dominant systematics on the \Bztopizpiz branching fraction arise
from the uncertainty on the \de resolution, the efficiency of the \piz
reconstruction, and the uncertainty on \B background event yields.
The significance of the \Bztopizpiz signal yield, taking systematic effects into account,
is $5.0\sigma$.
The systematic uncertainty on $C_{\piz\piz}$ is dominated
by the uncertainties
on the \B background asymmetry and tagging efficiency.

For \Bptohpiz the
dominant systematic uncertainties arise from the \fish signal PDF parameters, 
selection efficiencies, and  the \de resolution.
Additional systematics arise from uncertainties on
the \B background event yields and particle identification.
The systematic uncertainty on the charge asymmetries is 
dominated by the $1\%$ upper limit on the charge bias in the detector~\cite{chargeasym}.

\begin{figure}[!tbp]
\begin{center}
\includegraphics[width=0.62\linewidth]{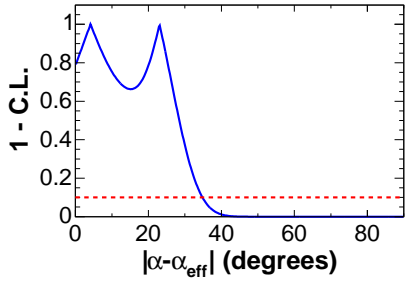}
\end{center}
\vspace{-0.5cm}
\caption{ 
 Constraints on the $|\deltapipi|$ in terms of confidence level.
 We find an upper bound  on $\left|\deltapipi\right|$  of $35^{\rm o}$ at the 90\% C.L.}
\label{fig:IsospinDalpha}
\end{figure}

To extract information on $\deltapipi$
we use the isospin relations~\cite{Isospin} 
in conjunction with \babar\ measurements of $C_{\pi^+\pi^-} = -0.09\pm  0.15\pm 0.04$~\cite{sin2alpha},
the branching fraction $\BR(\Bztopipi) = (4.7\pm 0.6\pm 0.2)\times 10^{-6}$~\cite{babarpipi},
the \Bztopizpiz and \Bptopipiz decay rates and
the $C_{\piz\piz}$ values reported here. 
We scan over all values of $\left| \deltapipi \right|$ 
and calculate a $\chi^2$ for the decay amplitudes 
using the method described in Ref.~\cite{CKMfitter}.
The $\chi^2$ is converted into a confidence level shown in Fig.~\ref{fig:IsospinDalpha},
from which we derive an upper bound on  $\left|\deltapipi\right|$  of $35^{\rm o}$ at the 90\% C.L.

In summary, we observe $61 \pm 17 \pm 5$
\Bztopizpiz events with a significance of $5.0\sigma$
including systematic uncertainties.
This corresponds to a branching fraction of $\BR(\Bztopizpiz) = ( 1.17 \pm 0.32 \pm
0.10 )\times 10^{-6}$, where the first error is statistical and the
second is systematic. 
We measure  the asymmetry
$C_{\piz\piz}= -0.12\pm 0.56\pm 0.06$. 
We report branching fractions  $\BR(\Bptopipiz) = (5.8\pm 0.6\pm 0.4)\times
10^{-6}$ and
$\BR(\Bptokpiz) = (12.0\pm 0.7 \pm 0.6)\times 10^{-6}$.
The charge asymmetries are $\acppipiz =  -0.01\pm 0.10\pm 0.02$
and $\acpkpiz = 0.06\pm 0.06\pm 0.01$;
we find no evidence for \CP violation. 
In contrast to the
recent measurements of charge asymmetry in \Btokppim decays~\cite{AKpi}, 
the \acpkpiz value reported here
is compatible with zero.
We use isospin relations on \Btopipi decay rates and asymmetries to find an upper  bound 
of $\left|\deltapipi\right|<35^{\rm o}$ at the 90\% C.L.

We are grateful for the excellent luminosity and machine conditions
provided by our \pep2\ colleagues, 
and for the substantial dedicated effort from
the computing organizations that support \babar.
The collaborating institutions wish to thank 
SLAC for its support and kind hospitality. 
This work is supported by
DOE
and NSF (USA),
NSERC (Canada),
IHEP (China),
CEA and
CNRS-IN2P3
(France),
BMBF and DFG
(Germany),
INFN (Italy),
FOM (The Netherlands),
NFR (Norway),
MIST (Russia), and
PPARC (United Kingdom). 
Individuals have received support from CONACyT (Mexico), A.~P.~Sloan Foundation, 
Research Corporation,
and Alexander von Humboldt Foundation.

\end{document}